\documentclass[11pt,a4paper]{article}
\pdfoutput=1 %
\usepackage{amsmath} 
\usepackage{jheppub}

\def\D{{\cal D}}
\def\g{\overline{\cal G}}
\def\gm{\gamma}

\title{Higgs boson production through $b \bar b$ annihilation at threshold in N$^3$LO QCD}

\author[a]{Taushif Ahmed,}
\author[a]{Narayan Rana}
\author[b]{and V. Ravindran}

\affiliation[a]{Regional Centre for Accelerator-based Particle Physics,\\ Harish-Chandra Research Institute, Allahabad, India}
\affiliation[b]{The Institute of Mathematical Sciences, Chennai, India }

\emailAdd{taushif@hri.res.in}
\emailAdd{narayan@hri.res.in}
\emailAdd{ravindra@imsc.res.in}

\abstract{We present threshold enhanced N$^3$LO QCD corrections to inclusive 
Higgs production through bottom anti-bottom annihilation at 
hadron colliders using threshold resummed cross section.  The resummed cross section   
is obtained using factorization properties and Sudakov resummation of 
the inclusive cross section.  We use the recent results on threshold N$^3$LO 
corrections in QCD for Drell-Yan production and three loop QCD corrections
to Higgs form factor with bottom anti-bottom quark to achieve this task.     
This is the first step towards the evaluation of complete N$^3$LO result.
We have numerically demonstrated the importance of such corrections at the LHC.} 

\preprint{HRI-RECAPP-2014-018}

\begin{document}
\allowdisplaybreaks[4]
\unitlength1cm
\maketitle
\flushbottom

\def\M{{\cal M}}
\def\ep{\epsilon}
\def\unM{\hat{\cal M}}
\def\unas{ \left( \frac{\hat{a}_s}{\mu_0^{\epsilon}} S_{\epsilon} \right) }
\def\rnM{{\cal M}}
\def\rnas{ \left( a_s  \right) }
\def\b0{\beta_0}
\def\cD{{\cal D}}
\def\cC{{\cal C}}
\def\ca{\text{\tiny C}_\text{\tiny A}}
\def\cf{\text{\tiny C}_\text{\tiny F}}

\def\ct{{\red []}}

\section{Introduction}
\setcounter{equation}{0}
\label{sec:intro}
The discovery of Higgs boson by ATLAS \cite{Aad:2012tfa} and CMS \cite{Chatrchyan:2012ufa}
collaborations of the Large Hadron Collider (LHC)
at CERN has not only shed the light on the dynamics behind the electroweak symmetry breaking
but also put the Standard Model (SM) of particle physics on a firmer ground.  In the SM, the 
elementary particles such as quarks, leptons and gauge bosons, $Z,W^\pm$ acquire 
their masses through spontaneous symmetry breaking (SSB).  
The Higgs mechanism provides the framework for 
SSB. The SM predicts the existence of a Higgs boson whose mass is a 
parameter of the model. 
The recent discovery of Higgs boson provides a valuable information on this, namely on
its mass which is about 125.5 GeV.   
The searches for the Higgs boson have been going on for several decades
in various experiments.  
Earlier experiments such as LEP \cite{higgslep} and Tevatron \cite{higgstev} played
an important role in the discovery by the LHC collaborations 
through narrowing down its possible mass range.
LEP excluded Higgs boson of mass below 114.4 GeV 
and their precision electroweak measurements \cite{lepprecis} 
hinted the mass less than 152 GeV at $95\%$ confidence level (CL), while
Tevatron excluded Higgs boson of mass in the range $162-166$ GeV at $95\%$ CL.   

Higgs bosons are produced dominantly at the LHC 
via gluon gluon fusion through top quark loop, while  
the sub-dominant ones 
are vector boson fusion, associated production of
Higgs boson with vector bosons, with top anti-top pairs and also in bottom anti-bottom 
annihilation.  The inclusive productions of Higgs boson in gluon gluon \cite{nnlo}, 
vector boson fusion processes \cite{bolzoni} and associated production with vector
bosons \cite{Han:1991ia} are known to next to next to leading order (NNLO) 
accuracy in QCD.     
Higgs production in bottom anti-bottom annihilation is also known to NNLO accuracy
in the variable flavour scheme (VFS) \cite{vfs, Harlander:2003ai}, while it is known to NLO in the fixed
flavour scheme (FFS) \cite{ffs}.   
In the minimal super symmetric standard model (MSSM), 
the coupling of bottom quarks to Higgs 
becomes large in the large $\tan\beta$ region, where
$\tan\beta$ is the ratio of vacuum expectation values of up and down type Higgs fields.
This can enhance contributions from bottom anti-bottom annihilation subprocesses.  

While the theoretical predictions of NNLO \cite{nnlo} and next to next to 
leading log (NNLL) \cite{nnll} QCD corrections and of two loop 
electroweak effects \cite{ewnnlo} played an important role in the Higgs discovery,
the theoretical uncertainties resulting from 
factorization and renormalization scales are not fully under control.  
Hence, the efforts to go beyond NNLO are going on intensively.
Some of the ingredients to obtain N$^3$LO QCD corrections
are already available.  For example, quark and gluon form
factors \cite{3lffmoch, Gehrmann:2005pd,Baikov:2009bg, Gehrmann:2010ue}, 
the mass factorization kernels \cite{Moch:2004pa}
and the renormalization constant \cite{Chetyrkin:1997un} for 
the effective operator describing the coupling of
Higgs boson with the SM fields in the infinite top quark mass 
limit up to three loop level in dimensional regularization are known for some time.  
In addition, NNLO soft contributions are known \cite{deFlorian:2012za} to all
orders in $\epsilon$ for both DY and Higgs productions using dimensional regularization
with space time dimension being $d=4+\epsilon$.
They were used to obtain the partial N$^3$LO
threshold effects \cite{n3losv, n3losvRavi} to 
Drell-Yan production of di-leptons and inclusive productions of Higgs boson through gluon gluon 
fusion and in bottom anti-bottom annihilation.  
Threshold contribution to the inclusive production
cross section is expanded in terms of $\delta(1-z)$ and ${\cal D}_i(z)$ where
\begin{eqnarray}
{\cal D}_i(z) = \left(\frac{\ln(1-z)}{1-z}\right)_+ 
\end{eqnarray}
with the scaling parameter 
$z=m_H^2/\hat s$ for Higgs 
and $z=m_{l^+l^-}^2/\hat s$ for DY.
Here $m_H$, $m_{l^+l^-}$ and $\hat s$ are mass of the Higgs boson,
invariant mass of the di-leptons and center of mass energy of the partonic reaction responsible for production mechanism respectively.
The missing $\delta(1-z)$ terms for the complete N$^3$LO threshold contributions 
to the Higgs production through gluon gluon fusion are now available 
due to the seminal work by Anastasiou et al \cite{Anastasiou:2014vaa} 
where the relevant soft contributions were obtained from the real radiations at 
N${}^3$LO level.  In \cite{Ahmed:2014cla}, we exploited the universal
structure of the soft radiations to obtain the corresponding soft
gluon contributions to DY production, which led to the evaluation of missing  
$\delta(1-z)$ part of the N$^3$LO threshold corrections, later
confirmed in \cite{Li:2014bfa}. 
For the Higgs production through $b \bar{b}$ annihilation,
till date, only partial N${}^3$LO threshold corrections are known \cite{n3losvRavi}.
It was not possible to determine the $\delta(1-z)$ at N$^3$LO
due to the lack of information on 
three loop finite part of bottom anti-bottom higgs form factor in QCD and the 
soft gluon radiation at N$^3$LO level.
The recent results on Higgs form factor with bottom anti-bottom by Gehrmann and Kara 
\cite{Gehrmann:2014vha} and on the universal soft distribution obtained
for the Drell-Yan production \cite{Ahmed:2014cla} can now be used   
to obtain $\delta(1-z)$ part of the threshold N$^3$LO contribution.
For the soft gluon radiations in the $b \bar{b}$ annihilation, 
the results from \cite{Ahmed:2014cla} can be 
used as they do not depend on the flavour of the incoming quark states.
We have set bottom quark mass to be zero throughout except in the Yukawa coupling. 

In the next section, we present the details of the threshold resummation and in section \ref{sec:res} we present 
our results for threshold N$^3$LO QCD contributions to Higgs production through $b\bar{b}$ annihilation at hadron colliders and their numerical impact .
The numerical impact of threshold enhanced  N$^3$LO contributions
is demonstrated for the LHC energy $\sqrt{s} = 14$ TeV by studying 
the stability of the perturbation theory under factorization and renormalization scales.
Finally we give a brief summary of our findings.

\section{Threshold resummation}
The interaction of bottom quarks and Higgs
boson is given by the action 
\begin{eqnarray}
S^b_{I} = - \lambda \int d^4 x \, \phi(x) \overline \psi_b(x) \psi_b(x)
\end{eqnarray}
where $\psi_b(x)$ denotes the bottom quark field and $\phi(x)$ the scalar field.
$\lambda$ is the Yukawa coupling given by $\sqrt{2} m_b/\nu$, with 
the bottom quark mass $m_b$ and 
the vacuum expectation value $\nu\approx 246$ GeV.  In MSSM,
for the pseudoscalar Higgs boson, we need to replace $\lambda \phi(x) \overline \psi_b(x) \psi_b(x)$ by 
$\tilde \lambda \tilde \phi(x) \overline \psi_b(x) \gamma_5 \psi_b(x)$ in the above
equation.  The MSSM couplings are given by
\[
 \tilde{\lambda} = \left\{
  \begin{array}{ll}
    -  \frac{\sqrt{2} m_b \sin\alpha}{\nu \cos\beta}  \,,& \qquad \tilde{\phi} = h\,,\\
    \phantom{-}  \frac{\sqrt{2} m_b \cos\alpha}{\nu \cos\beta}  \,,& \qquad \tilde{\phi}=H\,,\\
    \phantom{-}  \frac{\sqrt{2} m_b \tan\beta}{\nu } \,, & \qquad \tilde{\phi}=A\,
  \end{array}
  \right.
\]
respectively.  The angle $\alpha$ measures the mixing of weak and mass eigenstates
of neutral Higgs bosons.  We use VFS scheme throughout, hence except in the Yukawa coupling,
$m_b$ is taken to be zero like other light quarks in the theory.  

The inclusive Higgs production through bottom anti-bottom annihilation  
can be computed using
\begin{eqnarray}\label{sighad}
 \sigma^b(s,q^2) &=& \sigma^{(0)}_{b\overline b}(\mu_R^2) \sum_{a c} \int dx_1 dx_2 f_a (x_1,\mu_F^2) f_c (x_2,\mu_F^2) 
 {\Delta}^b_{a c} (\hat{s},q^2,\mu_F^2,\mu_R^2),
\end{eqnarray}
where $f_a(x_1,\mu_F^2)$ and $f_c(x_2,\mu_F^2)$ are parton distribution functions with
momentum fractions $x_1$ and $x_2$ respectively. 
$\mu_F$ is the factorization scale and $\hat{s} = x_1 x_2 s$ where 
$s$ ($\hat s$) is the square of hadronic (partonic) center of mass energy.  
The born cross section is given by
\begin{eqnarray}
\sigma^{(0)}_{b \overline b} = \frac{\pi \lambda^2(\mu_R^2)}{12 m_H^2} \, .
\end{eqnarray}
The born normalized partonic 
subprocesses after mass factorization are denoted by 
${\Delta}^b_{a c}$ where initial state partons are $a$ and $c$.  
Here, $q^2=m_H^2$, $\mu_F$ results from mass factorization and 
the renormalization scale $\mu_R$ is due to UV renormalization. 
${\Delta}^b_{a c}$ can be decomposed into two parts denoted by $\Delta^{sv}_b$ and
$\Delta^{b,R}_{ac}$.   
\begin{eqnarray}
\Delta^b_{ac} (z, q^2, \mu_R^2,\mu_F^2)
=\Delta^{sv}_b (z, q^2, \mu_R^2,\mu_F^2)
+\Delta^{b,R}_{ac} (z, q^2, \mu_R^2,\mu_F^2)
\end{eqnarray}
where $\Delta^{sv}_{b}$ contains only distributions such as $\delta(1-z)$ and
${\cal D}_i$, often called threshold contributions 
and the second term contains regular terms in $z$. 
Following \cite{n3losvRavi}, the threshold resumed cross section at the partonic level in $d=4+\epsilon$ dimensions is given by
\begin{equation}\label{sigma}
 \Delta^{sv}_{b} (z, q^2, \mu_R^2, \mu_F^2) = 
{\cal C} \exp \Big( \Psi^b (z, q^2, \mu_R^2, \mu_F^2, \epsilon )  \Big) \Big|_{\epsilon = 0}
\end{equation}
where the scaling variable $z=q^2/\hat s$
and $\Psi^b (z, q^2, \mu_R^2, \mu_F^2, \epsilon)$ is a finite distribution. 
The above resummed expression follows from factorization properties of the inclusive
cross section and Sudakov resummation of soft gluons in the QCD amplitudes.
The symbol ${\cal C}$ implies convolution with the following expansion 
\begin{equation}\label{conv}
 {\cal C} e^{f(z)} = \delta(1-z) + \frac{1}{1!} f(z) + \frac{1}{2!} f(z) \otimes f(z) + \ldots 
\end{equation}
Here $\otimes$ means Mellin convolution and $f(z)$ is a distribution of the kind $\delta(1-z)$ 
and ${\cal D}_i$.  We drop all the regular terms in $z$ in the evaluation of threshold contribution
$\Delta^{sv}_b$. 

In $d = 4 + \epsilon$ dimensions, the distribution $\Psi^b$ receives contributions from
UV renormalization constant $Z_b$ for Yukawa coupling $\lambda$,
the Higgs form factor $\hat F^b$ from bottom anti-bottom, 
soft gluon distribution $\Phi^b$ from the real radiations in partonic 
subprocesses and the mass factorization
kernels $\Gamma_{bb}$ that remove the collinear singularities 
from the initial state bottom quark states. That is
\begin{align}\label{psi}
\Psi^b (z, q^2, \mu_R^2, \mu_F^2, \epsilon) = &\Big( \ln \Big[ Z^b (\hat{a}_s, \mu_R^2, \mu^2, \epsilon) \Big]^2 
+ \ln \Big|  \hat{F}^b (\hat{a}_s, Q^2, \mu^2, \epsilon)  \Big|^2 \Big) \delta(1-z) \\
& + 2 \Phi^b (\hat{a}_s, q^2, \mu^2, z, \epsilon) - 2 {\cal C} \ln \Gamma_{bb} (\hat{a}_s, \mu^2, \mu_F^2, z, \epsilon) \, .
\end{align}
While the individual contributions are divergent, the sum is a finite distribution.
The scale $\mu$ is introduced to define the dimensionless coupling 
constant $\hat a_s=\hat g_s^2/16 \pi^2$ in dimensional regularization and $Q^2 = - q^2$.
The renormalized strong coupling constant $a_s(\mu_R^2)$ is related to bare $\hat a_s$ 
through strong coupling
constant renormalization $Z (\mu_R^2)$, 
\begin{eqnarray}
\hat a_s  = \Big( \frac{\mu}{\mu_R} \Big)^\epsilon 
Z (\mu_R^2) S_\epsilon^{-1} a_s(\mu_R^2), \quad \quad  S_\epsilon = \exp \Big( (\gamma_E-\ln 4 \pi) \frac{\epsilon}{2} \Big) \,.
\end{eqnarray}
where $\gamma_E$ is Euler-Mascheroni constant.
In $4+\epsilon$ dimensions, $Z(\mu_R^2)$ can be expressed in terms of the coefficients $\beta_i$ of the
$\beta$ function of the strong coupling RG equation. Solving the RGE, we obtain, up to 
three loop level   
\begin{eqnarray}
Z(\mu_R^2)= 1+ a_s(\mu_R^2) \frac{2 \beta_0 }{ \ep}
           + a_s^2(\mu_R^2) \Bigg(\frac{4 \beta_0^2 }{ \ep^2 }+
                  \frac{\beta_1 }{ \ep} \Bigg)
           + a_s^3(\mu_R^2) \Bigg( \frac{8 \beta_0^3 }{ \ep^3}
                   +\frac{14 \beta_0 \beta_1 }{ 3 \ep^2}
                   + \frac{2 \beta_2 }{ 3 \ep}\Bigg) .~~~~
\end{eqnarray}
The coefficients $\beta_0$, $\beta_1$ and $\beta_2$ are
\begin{align}
\beta_0&={11 \over 3 } C_A - {4 \over 3 } T_F n_f \, ,
\nonumber \\[0.5ex]
\beta_1&={34 \over 3 } C_A^2-4 T_F n_f C_F -{20 \over 3} T_F n_f C_A \, ,
\nonumber \\[0.5ex]
\beta_2&={2857 \over 54} C_A^3 
          -{1415 \over 27} C_A^2 T_F n_f
          +{158 \over 27} C_A T_F^2 n_f^2
\nonumber\\[0.5ex]
&          +{44 \over 9} C_F T_F^2 n_f^2
          -{205 \over 9} C_F C_A T_F n_f
          +2 C_F^2 T_F n_f 
\end{align}
where the $SU(N)$ QCD color factors are given by
\begin{equation}
C_A=N,\quad \quad \quad C_F={N^2-1 \over 2 N} , \quad \quad \quad
T_F={1 \over 2}
\end{equation}
and $n_f$ is the number of active flavours. \\
The UV renormalization constant $Z^b (\hat{a}_s, \mu_R^2, \mu^2, \epsilon)$ 
for Yukawa coupling $\lambda$ satisfies 
\begin{equation*}
 \mu_R^2 \frac{d}{d\mu_R^2} \ln Z^b (\hat{a}_s, \mu_R^2, \mu^2, \epsilon) =  
 \sum_{i=1}^{\infty}  a_s^i (\mu_R^2) \gamma^b_{i-1} \,.
\end{equation*}
The above renormalization group equation (RGE) can be solved in $4+\epsilon$ dimensions
to obtain up to ${\cal O}(a_s^3)$ level:
\begin{eqnarray}
Z^b(\hat a_s,\mu_R^2,\mu^2,\ep)& =&1
    + \hat a_s \left(\frac{\mu_R^2}{\mu^2}\right)^{\frac{\ep}{2}} S_{\ep}
       \Bigg[ \frac{1}{\ep}   \Bigg( 2~ \gamma^b_0 \Bigg)\Bigg]
   +\hat a_s^2 \left(\frac{\mu_R^2}{\mu^2}\right)^{{\ep }} S_{\ep}^2
       \Bigg[ \frac{1}{\ep^2}   \Bigg( 2~ \Big(\gamma^b_0\Big)^2 
           - 2~ \beta_0~ \gamma^b_0 \Bigg)
\nonumber\\[0.5ex]
&&       + \frac{1}{\ep}   \Bigg( \gamma^b_1 \Bigg)\Bigg]
    +\hat a_s^3 \left(\frac{\mu_R^2}{\mu^2}\right)^{3 \frac{\ep}{2}} S_{\ep}^3
       \Bigg[ \frac{1}{\ep^3}   \Bigg( \frac{4}{3}~ \Big(\gamma^b_0\Big)^3 
        - 4~ \beta_0~ \Big(\gamma^b_0\Big)^2 
          + \frac{8}{3}~ \beta_0^2~ \gamma^b_0 \Bigg)
\nonumber\\[0.5ex]
&&       + \frac{1}{\ep^2}   \Bigg( 2 ~\gamma^b_0 ~\gamma^b_1 
       - \frac{2}{3}~ \beta_1~ \gamma^b_0 
       - \frac{8}{3}~ \beta_0 ~\gamma^b_1 \Bigg)
       + \frac{1}{\ep}   \Bigg( \frac{2}{3} ~\gamma^b_2 \Bigg)\Bigg]
\end{eqnarray}
where the anomalous dimensions $\gamma^b_i$ can be obtained from
the quark mass anomalous dimensions \cite{vanRitbergen:1997va}
\begin{align}
\gamma^b_0&= 3 C_F \, ,
\nonumber \\[0.5ex]
\gamma^b_1&= \frac{3}{2} C_F^2
           + \frac{97}{6} C_F C_A
           - \frac{10}{3} C_F T_F n_f \, ,
\nonumber \\[0.5ex]
\gamma^b_2&= \frac{129}{2} C_F^3 
           - \frac{129}{4} C_F^2 C_A
           + \frac{11413}{108} C_F C_A^2
           +\Big(-46+48 \zeta_3\Big) C_F^2 T_F n_f
\nonumber \\[0.5ex]
&           +\left(-\frac{556}{27} -48 \zeta_3\right) C_F C_A T_F n_f
           - \frac{140}{27} C_F T_F^2 n_f^2 \, .
\end{align}
The bare form factor $\hat{F}^b (\hat{a}_s, Q^2, \mu^2, \epsilon)$ satisfies 
the following differential equation \cite{Sudakov:1954sw,Mueller:1979ih,Collins:1980ih,Sen:1981sd}
\begin{eqnarray}
 Q^2 \frac{d}{dQ^2} \ln \hat{F}^b  = \frac{1}{2} \Big[ K^b (\hat{a}_s, \frac{\mu_R^2}{\mu^2}, \epsilon ) + G^b (\hat{a}_s, \frac{Q^2}{\mu_R^2}, \frac{\mu_R^2}{\mu^2}, \epsilon ) \Big]
\end{eqnarray}
where $K^b$ contains all the poles in $\epsilon$ and $G^b$ contains the terms finite in $\epsilon$. Renormalization group invariance of  $\hat{F}^b (\hat{a}_s, Q^2, \mu^2, \epsilon)$ gives
\begin{eqnarray}
 \mu_R^2 \frac{d}{d\mu_R^2} K^b = - \mu_R^2 \frac{d}{d\mu_R^2} G^b 
= - \sum_{i=1}^{\infty}  a_s^i (\mu_R^2) A^q_i \, .
\end{eqnarray}
where $A^q_i$'s are the cusp anomalous dimensions \cite{cuspA, Moch:2004pa} given by
\begin{align}
 A^q_1 &= 4 C_F \,, 
\nonumber \\
 A^q_2 &= 8 C_F C_A \Big\{ \frac{67}{18} - \zeta_2 \Big\} + 8 C_F n_f \Big\{ -\frac{5}{9} \Big\} \,,
\nonumber \\
 A^q_3 &= 16 C_F C_A^2 \Big\{ \frac{245}{24} - \frac{67 \zeta_2 }{9} + \frac{11 \zeta_3}{6} 
                             + \frac{11 \zeta_2^2}{5} \Big\}
            + 16 C_F^2 n_f \Big\{ - \frac{55}{24} + 2 \zeta_3 \Big\}
\nonumber\\
&
            + 16 C_F C_A n_f \Big\{ - \frac{209}{108} + \frac{10 \zeta_2}{9} - \frac{7 \zeta_3}{3} \Big\}
            + 16 C_F n_f^2 \Big\{ - \frac{1}{27} \Big\} \,.
\end{align}
Expanding $\mu_R^2$ independent part 
of the solution of RG equation for $G^b$ as 
\begin{eqnarray}
G^b(a_s(Q^2), 1, \epsilon) = \sum_{i=1}^{\infty} a_s^i(Q^2) G^b_i(\epsilon),
\end{eqnarray}
one finds that $G^b_i$ can be decomposed in terms of  
collinear $B^q_i$ and soft $f^q_i$ anomalous dimensions as follows \cite{Ravindran:2004mb, Becher:2009cu, Gardi:2009qi}   
\begin{eqnarray}
 G^b_i (\epsilon) = 2 (B^q_i - \gamma^b_i) + f^q_i + C^b_i + \sum_{k=1}^{\infty} \epsilon^k g_i^{b,k} \,.
\end{eqnarray}
The collinear anomalous dimensions $B^q_i$ \cite{Moch:2004pa} are given by
\begin{align}
 B^q_1 &= 3 C_F \,,
 \nonumber \\
 B^q_2 &=  \frac{1}{2} \Big( C_F^2 \Big\{ 3 - 24 \zeta_2 + 48 \zeta_3 \Big\}
              + C_A C_F \Big\{ \frac{17}{3} + \frac{88}{3} \zeta_2 - 24 \zeta_3 \Big\}
                + n_f T_F C_F \Big\{ - \frac{4}{3} - \frac{32}{3} \zeta_2 \Big\} \Big) \,,
 \nonumber \\
 B^q_3 &= - 16 {C_A}^2 {C_F} \Big\{ \frac{1}{8} {\zeta_2}^2 - \frac{281}{27} {\zeta_2}
          + \frac{97}{9} {\zeta_3} - \frac{5}{2} {\zeta_5} + \frac{1657}{576} \Big\}
          + 16 {C_A} {C_F}^2 \Big\{ -\frac{247}{60} {\zeta_2}^2 + {\zeta_2} {\zeta_3} 
\nonumber\\
&
          - \frac{205}{24} {\zeta_2} +\frac{211}{12} {\zeta_3}
          + \frac{15}{2} {\zeta_5} + \frac{151}{64} \Big\}
+16 {C_A} {C_F} {n_f} \Big\{ \frac{1}{20} {\zeta_2}^2 - \frac{167}{54} {\zeta_2}
          + \frac{25}{18} {\zeta_3} + \frac{5}{4} \Big\}
\nonumber\\
&
+16 {C_F}^3 \Big\{ \frac{18}{5} {\zeta_2}^2 - 2 {\zeta_2} {\zeta_3}
          +\frac{9}{8} {\zeta_2} + \frac{17}{4} {\zeta_3} - 15 {\zeta_5} + \frac{29}{32} \Big\}
\nonumber\\
&
-16 {C_F}^2 {n_f} \Big\{ - \frac{29}{30} {\zeta_2}^2 - \frac{5}{12} {\zeta_2}
          +\frac{17}{6} {\zeta_3} + \frac{23}{16} \Big\}
-16 {C_F} {n_f}^2 \Big\{ -\frac{5}{27} {\zeta_2} + \frac{1}{9} {\zeta_3} + \frac{17}{144}\Big\} \,,
 \nonumber 
\end{align}
the soft anomalous dimensions $f_i^q$ \cite{Ravindran:2004mb} are 
\begin{align}
 f_1^q &= 0 \,,
\nonumber \\
 f_2^q &= C_A C_F \Big\{ -\frac{22}{3} {\zeta_2} - 28 {\zeta_3} + \frac{808}{27} \Big\}
        + C_F n_f T_F \Big\{ \frac{8}{3} {\zeta_2} - \frac{224}{27} \Big\} \,,
\nonumber \\
 f_3^q &= {C_A}^2 C_F \Big\{ \frac{352}{5} {\zeta_2}^2 + \frac{176}{3} {\zeta_2} {\zeta_3}
        - \frac{12650}{81} {\zeta_2} - \frac{1316}{3} {\zeta_3} + 192 {\zeta_5}
        + \frac{136781}{729}\Big\}
\nonumber \\
&
        + {C_A} {C_F} {n_f} \Big\{ - \frac{96}{5} {\zeta_2}^2 
        + \frac{2828}{81} {\zeta_2}
        + \frac{728}{27} {\zeta_3} - \frac{11842}{729} \Big\} 
\nonumber \\
&
        + {C_F}^2 {n_f} \Big\{ \frac{32}{5} {\zeta_2}^2 + 4 {\zeta_2} 
        + \frac{304}{9} {\zeta_3} - \frac{1711}{27} \Big\}
        + {C_F} {n_f}^2 \Big\{ - \frac{40}{27} {\zeta_2} + \frac{112}{27} {\zeta_3}
        - \frac{2080}{729} \Big\}
\end{align}
and the constants $C^b_i$ are given by 
\begin{equation}
C^b_1 = 0,\quad \quad  C^b_2 = - 2 \beta_0 g_1^{b,1}, \quad \quad
C^b_3 = - 2 \beta_1 g_1^{b,1} - 2 \beta_0 ( g_2^{b,1} + 2 \beta_0  g_1^{b,2}). 
\end{equation}
Since $G^b_1 (\epsilon)$ and $G^b_2 (\epsilon)$ are
known to all orders in $\epsilon$ and $G^b_3 (\epsilon)$ is known to ${\cal O} (\epsilon^3)$ 
\cite{Gehrmann:2014vha},
the coefficients $g_i^{b,k}$ for $i=1,2,3$ can be readily obtained from $G^b_i(\epsilon)$. 
The relevant one loop terms are found to be 
\begin{align}
 g_1^{b,1} = C_F ( - 2 + \zeta_2 ),\, \quad \quad 
 g_1^{b,2} = C_F ( 2 - \frac{7}{3} \zeta_3 ) , \,\quad \quad 
 g_1^{b,3} = C_F ( - 2 + \frac{1}{4} \zeta_2 + \frac{47}{80} \zeta_2^2 )\,,
\end{align}
the relevant two loop terms \cite{Harlander:2003ai, n3losvRavi} are
\begin{align}
 \nonumber \\
g_2^{b,1} &= C_F n_f \Big\{ \frac{616}{81} + \frac{10}{9} \zeta_2 - \frac{8}{3} \zeta_3 \Big\}
          + C_F C_A \Big\{ - \frac{2122}{81} - \frac{103}{9} \zeta_2 + \frac{88}{5} {\zeta_2}^2 + \frac{152}{3} \zeta_3 \Big\}
\nonumber \\
          &~~ + C_F^2 \Big\{ 8 + 32 \zeta_2 - \frac{88}{5} {\zeta_2}^2 - 60 \zeta_3 \Big\}  \,,
 \nonumber \\
 g_2^{b,2} &=
C_F n_f \Bigg\{ \frac{7}{12} {\zeta_2}^2 - \frac{55}{27} \zeta_2 + \frac{130}{27} \zeta_3 - \frac{3100}{243} \Bigg\}
+ C_A C_F  \Bigg\{ - \frac{365}{24} {\zeta_2}^2 + \frac{89}{3} \zeta_2 \zeta_3 + \frac{1079}{54} \zeta_2 
\nonumber \\
&~~ - \frac{2923}{27} \zeta_3 - 51 \zeta_5 + \frac{9142}{243} \Bigg\}
+ C_F^2 \Bigg\{ \frac{ 96}{5} {\zeta_2}^2 - 28 \zeta_2 \zeta_3 
 - 44 \zeta_2 + 116 \zeta_3 + 12 \zeta_5 - 24 \Bigg\}
 \nonumber
\end{align}
and finally the relevant three loop term \cite{Gehrmann:2014vha} is  
\begin{align}
  g_3^{b,1} &= 
C_A^2 C_F   \Big\{ - \frac{6152}{63} {\zeta_2}^3 + \frac{2738}{9} {\zeta_2}^2
 + \frac{976}{9} \zeta_2 \zeta_3 - \frac{342263}{486} \zeta_2
 - \frac{1136}{3} {\zeta_3}^2 + \frac{19582}{9} \zeta_3 
\nonumber \\
&
 + \frac{1228}{3} \zeta_5 
 + \frac{4095263}{8748} \Big\}
+ C_A C_F^2  \Bigg\{ - \frac{15448}{105} {\zeta_2}^3 - \frac{3634}{45} {\zeta_2}^2
 - \frac{2584}{3} \zeta_2 \zeta_3 + \frac{13357}{9} \zeta_2 
\nonumber \\
&
 + 296 \zeta_3^2
 - \frac{11570}{9} \zeta_3 - \frac{1940}{3} \zeta_5 - \frac{613}{3} \Bigg\}
+ C_A C_F n_f  \Bigg\{ - \frac{1064}{45} {\zeta_2}^2 + \frac{392}{9} \zeta_2 \zeta_3 
 + \frac{44551}{243} \zeta_2 
\nonumber \\
&
 - \frac{41552}{81} \zeta_3 
 - 72 \zeta_5 - \frac{6119}{4374} \Bigg\}
+ C_F^2 n_f  \Bigg\{ \frac{772}{45} {\zeta_2}^2 - \frac{152}{3} \zeta_2 \zeta_3
  - \frac{3173}{18} \zeta_2 + \frac{15956}{27} \zeta_3 -\frac{368}{3} \zeta_5
\nonumber \\
&  
  + \frac{32899}{324}\Bigg\}
+ C_F n_f^2  \Bigg\{ - \frac{40}{9} {\zeta_2}^2 - \frac{892}{81} \zeta_2 
  + \frac{320}{81} \zeta_3 - \frac{27352}{2187} \Bigg\}
+ C_F^3 \Bigg\{ \frac{21584}{105} {\zeta_2}^3 - \frac{1644}{5} {\zeta_2}^2
\nonumber \\
&
  + 624 \zeta_2 \zeta_3 
  - 275 \zeta_2 + 48 \zeta_3^2 
  - 2142 \zeta_3 + 1272 \zeta_5 + 603 \Bigg\} \,.
\end{align}
The mass factorization kernel $\Gamma_{bb}(z,\mu_F^2,\epsilon )$ removes the collinear singularities resulting from massless partons.  It satisfies  
the following RG equation 
\begin{equation}
 \mu_F^2 \frac{d}{d\mu_F^2} \Gamma_{bb}(z,\mu_F^2,\epsilon) = \frac{1}{2} \sum_c P_{bc} \left(z,\mu_F^2\right) \otimes \Gamma_{cb} \left(z,\mu_F^2,\epsilon \right) \, ,
\end{equation}
where $P_{bc} \left(z,\mu_F^2\right)$ are Altarelli-Parisi splitting functions.  In perturbative QCD,  
we can expand them as
\begin{eqnarray}
P_{bc}(z,\mu_F^2)=\sum_{i=1}^\infty a_s^i(\mu_F^2) P^{(i-1)}_{bc}(z).
\end{eqnarray}  
The off diagonal splitting functions are regular as $z\rightarrow 1$. On the other hand
diagonal ones contain $\delta(1-z)$ and ${\cal D}_0$ as well as regular terms i.e.
\begin{equation}
P^{(i)}_{bb}(z) = 2 (B_{i+1}^b \delta(1-z)+ A_{i+1}^b {\cal D}_0) + P_{reg,bb}^{(i)}(z) \,.
\end{equation}
We find that the regular part of the splitting function, $P_{reg,bb}^{(i)}$, does not
contribute to threshold corrections. \\
The fact that $\Delta_b^{sv}$ is finite as $\epsilon \rightarrow 0$ implies 
that soft distribution function 
$\Phi^{b}(\hat{a}_s, q^2, \mu^2, z, \epsilon)$ also 
satisfies Sudakov type differential equations \cite{n3losvRavi} namely
\begin{equation*}
 q^2 \frac{d}{dq^2} \Phi^b  = \frac{1}{2} \Big[ \overline K^b (\hat{a}_s, \frac{\mu_R^2}{\mu^2}, z,
\epsilon ) + \overline G^b (\hat{a}_s, \frac{q^2}{\mu_R^2}, \frac{\mu_R^2}{\mu^2}, z, \epsilon ) \Big] \, .
\end{equation*}
$\overline{K}^b$ and $\overline{G}^b$ take the forms similar to those of $K^b$ and $G^b$ of the
form factors.  This guarantees $\Psi^b$ is finite order by order in perturbation theory 
as $\epsilon \rightarrow 0$.  
The solution to the above equation is found to be
\begin{eqnarray}
\Phi^b = \sum_{i=1}^\infty {\hat a}_s^i  \left({q^2 (1-z)^2 \over \mu^2}\right)^{i {\epsilon \over 2}}
S_\epsilon^i \left({i \epsilon \over 1-z}\right) \hat \phi^{b,(i)} (\epsilon)
\end{eqnarray}
where 
\begin{eqnarray}
\hat \phi^{b,(i)}(\epsilon) = {1 \over i \epsilon}\big[\overline K^{b,(i)}(\epsilon) 
+ \overline G^{b,(i)}(\epsilon)\big]   
\end{eqnarray}
and the constants $\overline K^b$ can be obtained using 
\begin{eqnarray}
\mu_R^2 \frac{d \overline K^b}{d\mu_R^2} = - \delta(1-z) \mu_R^2 \frac{d K^b }{ d\mu_R^2} \,.
\end{eqnarray}
This implies that $\overline K^{b,(i)} (\epsilon)$ can be 
written in terms of $A^q_i$ and $\beta_i$.  
Defining $\overline {\cal G}^b_i(\epsilon)$ through
\begin{eqnarray}
\sum_{i=1}^\infty \hat a_s^i \left( {q_z^2  \over \mu^2}\right)^{i {\epsilon \over 2}}
S_\epsilon^i \overline G^{b,(i)}(\epsilon) =
\sum_{i=1}^\infty a_s^i(q_z^2 ) 
\overline {\cal G}^b_i(\epsilon)
\end{eqnarray}
where $q_z^2=q^2 (1-z)^2$ and using the fact that $\Delta^{sv}_b$ is finite as $\epsilon \rightarrow 0$, we can express
$\overline {\cal G}^b_i(\epsilon)$ as 
\begin{align}
\overline {\cal G}^{~b}_{1}(\ep)&= - f_1^q + \sum_{k=1}^\infty \ep^k \overline {\cal G}^{~b,(k)}_{1} \,,
\nonumber \\
\overline {\cal G}^{~b}_{2}(\ep)&= - f_2^q - 2 \beta_0 \overline{\cal G}_{1}^{~b,(1)}
                                   + \sum_{k=1}^\infty\ep^k  \overline {\cal G}^{~b,(k)}_{2} \,,
\nonumber \\
\overline {\cal G}^{~b}_{3}(\ep)&= - f_3^q - 2 \beta_1 \overline{\cal G}_{1}^{~b,(1)}
                                   - 2 \beta_0 \left(\overline{\cal G}_{2}^{~b,(1)}
                                   + 2 \beta_0 \overline{\cal G}_{1}^{~b,(2)}\right)
                                   + \sum_{k=1}^\infty \ep^k \overline {\cal G}^{~b,(k)}_{3} \,.
\label{OGI}
\end{align}
The constants $\overline{\cal G}^{b,j}_i$ arise from the soft part of the partonic 
reactions.  Since the soft part does not depend on the hard process, 
$\overline{\cal G}^{b}_i (\epsilon)$ can be directly obtained from  
$\overline{\cal G}^{q}_i (\epsilon)$ that contributes to Drell-Yan production  
\cite{n3losvRavi}
\begin{align}
\overline {\cal G}^b_i (\epsilon)= \overline {\cal G}^q_i(\epsilon) \,.
\label{mna}
\end{align}
We list the relevant ones that contribute up to N$^3$LO level. The terms required for one and two loops \cite{n3losvRavi} are 
\begin{align*}
  {\overline {\cal G}}^{q,(1)}_1 &= C_F ( - 3 \zeta_2 ) \,, \\
  {\overline {\cal G}}^{q,(2)}_1 &= C_F ( \frac{7}{3} \zeta_3 ) \,, \\
  {\overline {\cal G}}^{q,(3)}_1 &=  C_F ( - \frac{3}{16} {\zeta_2}^2 ) \,, \\
  {\overline {\cal G}}^{q,(1)}_2 &=  C_F n_f  \Big( - \frac{328}{81} + \frac{70}{9} \zeta_2 + \frac{32}{3} \zeta_3 \Big)
             + C_A C_F  \Big( \frac{2428}{81} - \frac{469}{9} \zeta_2 
                       + 4 {\zeta_2}^2 - \frac{176}{3} \zeta_3 \Big) \,, \\
  {\overline {\cal G}}^{q,(2)}_2 &=  C_A C_F \Big( \frac{11}{40} {\zeta_2}^2 - \frac{203}{3} {\zeta_2} {\zeta_3}
             + \frac{1414}{27} {\zeta_2} + \frac{2077}{27} {\zeta_3} + 43 {\zeta_5} - \frac{7288}{243}  \Big) \\
           & + C_F n_f \Big( -\frac{1}{20} {\zeta_2}^2 - \frac{196}{27} {\zeta_2} - \frac{310}{27} {\zeta_3} + \frac{976}{243} \Big) \\
\end{align*}
and for three loops \cite{Ahmed:2014cla} 
\begin{align*}
%
{\overline {\cal G}}^{q,(1)}_3 &= 
C_F \Big\{  {C_A}^2 \Big(\frac{152}{63} \;{\zeta_2}^3 + \frac{1964}{9} \;{\zeta_2}^2
+ \frac{11000}{9} \;{\zeta_2} {\zeta_3} - \frac{765127}{486} \;{\zeta_2}
+\frac{536}{3} \;{\zeta_3}^2 - \frac{59648}{27} \;{\zeta_3} 
\\
&
- \frac{1430}{3} \;{\zeta_5}
+\frac{7135981}{8748}\Big)
+{C_A} {n_f} \
\Big(-\frac{532}{9} \;{\zeta_2}^2 - \frac{1208}{9} \;{\zeta_2} {\zeta_3}
+\frac{105059}{243} \;{\zeta_2} + \frac{45956}{81} \;{\zeta_3} 
\\
&
+\frac{148}{3} \;{\zeta_5} - \frac{716509}{4374} \Big)
+ {C_F} {n_f} \
\Big(\frac{152}{15} \;{\zeta_2}^2 
- 88 \;{\zeta_2} {\zeta_3} 
+\frac{605}{6} \;{\zeta_2} + \frac{2536}{27} \;{\zeta_3}
+\frac{112}{3} \;{\zeta_5} 
\\
&
- \frac{42727}{324}\Big)
+ {n_f}^2 \
\Big(\frac{32}{9} \;{\zeta_2}^2 - \frac{1996}{81} \;{\zeta_2}
-\frac{2720}{81} \;{\zeta_3} + \frac{11584}{2187}\Big)  \Big\} \,.
\end{align*}
%
%
%
\section{Results}
\label{sec:res}
Expanding eqn.(\ref{sigma}) in powers of $a_s(\mu_R^2)$ using eqn(\ref{conv}) 
and performing the convolutions we find
\begin{eqnarray}
\Delta_b^{sv}(z) = 
\sum_{i=0}^\infty a_s^i(\mu_R^2) \Delta_{b}^{sv,(i)} (z,\mu_R^2) 
\end{eqnarray}
where $\Delta_{b}^{sv,(i)}$ can be expressed in terms of the distributions 
$\delta(1-z)$ and ${\cal D}_i$:
\begin{eqnarray}
\Delta_{b}^{sv,(i)} =
\Delta_{b}^{sv,(i)} (\mu_R^2)|_\delta
\delta(1-z) 
+ \sum_{j=0}^{2i-1} 
\Delta_{b}^{sv,(i)} (\mu_R^2)|_{{\cal D}_j}
{\cal D}_j \, .
\end{eqnarray}
%
%
The results up to N$^3$LO are given below
\begin{align}
\Delta_{b}^{sv,(1)} &= \Big\{
        \delta(1-z) \Big(   2 \g^{b,(1)}_{1} + 2 g^{b,1}_{1}  + 3 \zeta_2  A^b_1   \Big)
       - 2  \D_0  f^b_1 
       + 4 \D_1 A^b_1 
         \Big\} \,,
\nonumber \\
%
%
 \Delta_{b}^{sv,(2)} &= \Big\{
    \delta(1-z) \Big(
         \g^{b,(1)}_{2} + 2 {\g^{b,(1)}_{1}}^2 + g^{b,1}_{2} + 4 g^{b,1}_{1} \g^{b,(1)}_{1} + 2 {g^{b,1}_{1}}^2 
       + \beta_0 \Big(
            2 \g^{b,(2)}_{1}
          + 2 g^{b,2}_{1} 
          \Big)
\nonumber \\
&
       - 8  \zeta_3 f^b_1 A^b_1 
       + \zeta_2  \Big(
            3 A^b_2
          - 2 \Big(f^b_1\Big)^2   
          + 6 \g^{b,(1)}_{1} A^b_1
          + 6 g^{b,1}_{1} A^b_1
          \Big)
       + \zeta_2 \beta_0 \Big(
          - 6 \gm_0^b
          + 6 B_1^b
\nonumber \\
&
          + 3  f^b_1
          \Big)
          + \frac{37}{10}  \zeta_2^2  \Big( A^b_1 \Big)^2
     \Big)
 + \D_0 \Big( 
       - 4 \beta_0  \g^{b,(1)}_{1}
       + 16 \zeta_3  \Big( A^b_1 \Big)^2
       + 2  \zeta_2   f^b_1  A^b_1    
          - 2  f^b_2 
\nonumber \\
&
          - 4 \g^{b,(1)}_{1}  f^b_1
          - 4 g^{b,1}_{1}  f^b_1 
    \Big)
 + \D_1  \Big(
         4 \beta_0  f^b_1
       - 4  \zeta_2   \Big( A^b_1 \Big)^2
       +  4  A^b_2
       + 4 \Big( f^b_1 \Big)^2
       + 8 \g^{b,(1)}_{1}  A^b_1
\nonumber \\
&
       + 8 g^{b,1}_{1}  A^b_1
     \Big)  
     + \D_2  \Big(
          - 4 \beta_0  A^b_1 
          - 12 f^b_1  A^b_1  
      \Big)
       + 8  \D_3   \Big( A^b_1 \Big)^2   
    \Big\}  \,,
\nonumber \\
%
%
 \Delta_{b}^{sv,(3)} &=  
    \delta(1-z)  \Big\{  
            \frac{2}{3} \g^{b,(1)}_{3}
          + 2 \g^{b,(1)}_{1} \g^{b,(1)}_{2}
          + \frac{4}{3} {\g^{b,(1)}_{1}}^3
          + \frac{2}{3} g^{b,1}_{3}
          + 2 g^{b,1}_{2}  \g^{b,(1)}_{1}
          + 2 g^{b,1}_{1}  \g^{b,(1)}_{2}
\nonumber \\
&
          + 4 g^{b,1}_{1} {\g^{b,(1)}_{1}}^2
          + 2 g^{b,1}_{1}  g^{b,1}_{2}
          + 4 {g^{b,1}_{1}}^2  \g^{b,(1)}_{1}   
          + \frac{4}{3} {g^{b,1}_{1}}^3
       + \frac{4}{3} \beta_1  \Big(
            \g^{b,(2)}_{1}
          + g^{b,2}_{1}
          \Big)
       + 4 \beta_0  \Big(
            \frac{1}{3} \g^{b,(2)}_{2}
\nonumber \\
&
          +  \g^{b,(1)}_{1}  \g^{b,(2)}_{1}
          + \frac{1}{3} g^{b,2}_{2}
          +  g^{b,2}_{1}  \g^{b,(1)}_{1}
          +  g^{b,1}_{1}  \g^{b,(2)}_{1}
          +  g^{b,1}_{1}  g^{b,2}_{1}
          \Big)
       + \frac{8}{3} \beta_0^2  \Big(
            \g^{b,(3)}_{1}
          + g^{b,3}_{1}
          \Big)             
\nonumber \\
&
       - 32 \zeta_5 \Big( 3f^b_1 + 2 \beta_0  \Big) \Big( A^b_1 \Big)^2 
       - 8 \zeta_3  \Big(
            f^b_2  A^b_1
          +  f^b_1  A^b_2 
          + \frac{1}{3} \Big( f^b_1 \Big)^3
          + 2 \g^{b,(1)}_{1}  f^b_1  A^b_1 
\nonumber \\
&
          + 2 g^{b,1}_{1}  f^b_1  A^b_1 
          \Big)
       - 8 \zeta_3 \beta_0  \Big(
           \Big( f^b_1 \Big)^2    
          + 2 \g^{b,(1)}_{1} A^b_1 
          \Big)  
          + \frac{160}{3} \zeta_3^2  \Big(A^b_1\Big)^3
       + \zeta_2  \Big(
            3  A^b_3 
          - 4  f^b_1   f^b_2
\nonumber \\
&
          + 3 \g^{b,(1)}_{2}  A^b_1 
          + 6 \g^{b,(1)}_{1}  A^b_2
          - 4 \g^{b,(1)}_{1} \Big( f^b_1 \Big)^2
          + 6 {\g^{b,(1)}_{1}}^2  A^b_1 
          + 3 g^{b,1}_{2}  A^b_1 
          + 6 g^{b,1}_{1}  A^b_2   
\nonumber \\
&
          - 4 g^{b,1}_{1}  \Big( f^b_1 \Big)^2 
          + 12 g^{b,1}_{1}  \g^{b,(1)}_{1}  A^b_1 
          + 6 {g^{b,1}_{1}}^2  A^b_1
          \Big)
       - 3 \zeta_2 \beta_1  \Big(
            2 \gm_0^b
          - 2 B_1^b
          -   f^b_1 
          \Big)
\nonumber \\
&
       + 6 \zeta_2 \beta_0  \Big(
          - 2 \gm_1^b
          + 2 B_2^b
          +   f^b_2 
          +  \g^{b,(2)}_{1}  A^b_1  
          - 2 \g^{b,(1)}_{1} \gm_0^b
          + 2 \g^{b,(1)}_{1} B_1^b   
          - \frac{1}{3} \g^{b,(1)}_{1}  f^b_1 
          +  g^{b,2}_{1}  A^b_1 
\nonumber \\
&
          - 2 g^{b,1}_{1} \gm_0^b
          + 2 g^{b,1}_{1} B_1^b
          +  g^{b,1}_{1}  f^b_1 
          \Big)
       - 12 \zeta_2 \beta_0^2   g^{b,1}_{1}
       + 40 \zeta_2 \zeta_3   f^b_1  \Big( A^b_1 \Big)^2  
       + 32  \zeta_2 \zeta_3 \beta_0   \Big( A^b_1 \Big)^2
\nonumber \\
&
       + \frac{37}{5} \zeta_2^2  \Big(
             A^b_1  A^b_2  
          - \frac{38}{37}  \Big( f^b_1 \Big)^2  A^b_1
          +  \g^{b,(1)}_{1}  \Big( A^b_1 \Big)^2
          +  g^{b,1}_{1} \Big( A^b_1 \Big)^2
          \Big)
       + \zeta_2^2 \beta_0  \Big(
           18  A^b_1 B_1^b 
\nonumber \\
&  
          - 18  A^b_1 \gm_0^b
          + f^b_1  A^b_1 
          \Big)
       - 3 \zeta_2^2 \beta_0^2   A^b_1 
       - \frac{283}{42}  \zeta_2^3  \Big( A^b_1 \Big)^3
     \Big\}  
    + \D_0   \Big\{
          - 4 \beta_1  \g^{b,(1)}_{1}
       - 4 \beta_0  \Big(
             \g^{b,(1)}_{2}
\nonumber \\
&  
          +  \g^{b,(2)}_{1}   f^b_1 
          + 2 {\g^{b,(1)}_{1}}^2
          +   g^{b,2}_{1}  f^b_1 
          + 2 g^{b,1}_{1} \g^{b,(1)}_{1}
          \Big)
       - 8 \beta_0^2  \g^{b,(2)}_{1}
       + 192 \zeta_5  \Big( A^b_1 \Big)^3
       + 32 \zeta_3  \Big(
            A^b_1  A^b_2  
\nonumber \\
&
          + \Big( f^b_1 \Big)^2  A^b_1 
          + \g^{b,(1)}_{1}  \Big( A^b_1 \Big)^2
          + g^{b,1}_{1}  \Big( A^b_1 \Big)^2
          \Big)
       + 48  \zeta_3 \beta_0  f^b_1 A^b_1 
       + 2 \zeta_2  \Big(
             f^b_2  A^b_1 
          +  f^b_1  A^b_2 
\nonumber \\
&
          + 2 \Big( f^b_1 \Big)^3
          + 2 \g^{b,(1)}_{1}  f^b_1  A^b_1 
          + 2 g^{b,1}_{1}  f^b_1  A^b_1 
          \Big)   
          + 2 \zeta_2 \beta_0  \Big(
            6 f^b_1  \gm_0^b
          - 6 f^b_1  B_1^b
          +  \Big( f^b_1 \Big)^2
\nonumber \\
&
          + 2 \g^{b,(1)}_{1}  A^b_1 
          \Big)
       - 80 \zeta_2 \zeta_3  \Big( A^b_1 \Big)^3
       + 23  \zeta_2^2  f^b_1  \Big( A^b_1 \Big)^2
       + 16 \zeta_2^2 \beta_0  \Big( A^b_1 \Big)^2
       - 2  \Big(
             f^b_3 
          +  \g^{b,(1)}_{2}  f^b_1 
\nonumber \\
&
          + 2 \g^{b,(1)}_{1}  f^b_2 
          + 2 {\g^{b,(1)}_{1}}^2  f^b_1 
          +   g^{b,1}_{2}  f^b_1
          + 2 g^{b,1}_{1}  f^b_2 
          + 4 g^{b,1}_{1} \g^{b,(1)}_{1}  f^b_1 
          + 2 {g^{b,1}_{1}}^2  f^b_1
          \Big)
     \Big\}  
       + \D_1   \Big\{  
          4 \beta_1  f^b_1 
\nonumber \\
&
       + 8 \beta_0  \Big(
            f^b_2
          +  \g^{b,(2)}_{1}  A^b_1  
          + 3 \g^{b,(1)}_{1} f^b_1  
          +  g^{b,2}_{1}  A^b_1 
          +  g^{b,1}_{1} f^b_1 
          \Big)
       + 16 \beta_0^2   \g^{b,(1)}_{1}
       - 160  \zeta_3  f^b_1   \Big( A^b_1 \Big)^2
\nonumber \\
&
       - 96  \zeta_3 \beta_0  \Big( A^b_1 \Big)^2
       - 8 \zeta_2  \Big(
           A^b_1  A^b_2  
          + \frac{7}{2} \Big(f^b_1\Big)^2   A^b_1 
          + \g^{b,(1)}_{1} \Big( A^b_1 \Big)^2     
          +  g^{b,1}_{1}  \Big( A^b_1 \Big)^2
          \Big)
\nonumber \\
&
       - 24 \zeta_2 \beta_0  \Big(
            A^b_1  \gm_0^b
          -  A^b_1  B_1^b
          +  f^b_1  A^b_1
          \Big)
       - 46  \zeta_2^2    \Big( A^b_1 \Big)^3
       + 4 \Big(
              A^b_3 
          + 2 f^b_1   f^b_2 
          +  \g^{b,(1)}_{2}  A^b_1 
\nonumber \\
&
          + 2 \g^{b,(1)}_{1} A^b_2   
          + 2 \g^{b,(1)}_{1}  \Big( f^b_1 \Big)^2  
          + 2 {\g^{b,(1)}_{1}}^2 A^b_1
          +  g^{b,1}_{2}   A^b_1 
          + 2 g^{b,1}_{1}  A^b_2 
          + 2 g^{b,1}_{1}  \Big( f^b_1 \Big)^2
\nonumber \\
&
          + 4 g^{b,1}_{1} \g^{b,(1)}_{1} A^b_1 
          + 2 {g^{b,1}_{1}}^2   A^b_1 
          \Big)
     \Big\}
       + \D_2 \Big\{
         - 4 \beta_1  A^b_1 
       - 8 \beta_0  \Big(
           A^b_2 
          + \frac{3}{2} \Big( f^b_1 \Big)^2   
          + 4 \g^{b,(1)}_{1}  A^b_1 
\nonumber \\
&
          +  g^{b,1}_{1} A^b_1 
          \Big)
       - 8  \beta_0^2  
           f^b_1 
       + 160 \zeta_3   \Big( A^b_1 \Big)^3
       + 60 \zeta_2   f^b_1 \Big( A^b_1 \Big)^2
       + 36 \zeta_2 \beta_0    \Big( A^b_1 \Big)^2
       - 12  \Big(
           f^b_2  A^b_1 
\nonumber \\
&
          +  f^b_1  A^b_2 
          + \frac{1}{3} \Big( f^b_1 \Big)^3  
          + 2 \g^{b,(1)}_{1} f^b_1  A^b_1 
          + 2 g^{b,1}_{1}   f^b_1  A^b_1 
          )
     \Big\}
       + \D_3   \Big\{ 
         \frac{80}{3} \beta_0  f^b_1  A^b_1 
         + \frac{16}{3}\beta_0^2   A^b_1
\nonumber \\
&
       - 40 \zeta_2  \Big( A^b_1 \Big)^3
       + 16 \Big(
            A^b_1  A^b_2 
          + \Big( f^b_1 \Big)^2  A^b_1 
          + \g^{b,(1)}_{1}\Big( A^b_1 \Big)^2  
          + g^{b,1}_{1} \Big( A^b_1 \Big)^2
          \Big)
     \Big\}
\nonumber \\ \label{ci3}
&
       + \D_4  \Big\{
          - \frac{40}{3} \beta_0     \Big( A^b_1 \Big)^2
       - 20  f^b_1 \Big( A^b_1 \Big)^2
     \Big\}          
       + 8 \D_5 \Big( A^b_1 \Big)^3  \,.
\end{align}
%
%
In the above equation $A_i^b=A_i^q,~ B_i^b=B_i^q$, $f_i^b=f_i^q$ and we have set $\mu_R^2=\mu_F^2=q^2$.

The finite $\Delta_{b}^{sv,(i)}(Q^2)$ depend on the anomalous dimensions 
$A^b_i$, $B^b_i$, $f^b_i$ and $\gamma^b_i$, the $\beta$ function coefficients
$\beta_i$ and $\epsilon$ expansion coefficients of $G^b(\epsilon)$, $g^{b,i}_j$s and
of the corresponding $\overline {\cal G}^b(\epsilon)$, $\overline{\cal G}^{b,i}_j$s.  
The results for the $\Delta_{b}^{sv,(1)}$ and $\Delta_{b}^{sv,(2)}$ agree 
with those from the exact NLO and NNLO results \cite{Harlander:2003ai}. 
At N$^3$LO level, only $\Delta_{b}^{sv,(3)} |_{{\cal D}_i}$'s were known \cite{n3losvRavi} as 
the terms $g^{b,2}_2,g^{b,1}_3$ of the form factor and $\overline {\cal G}^{q,(2)}_2$,
$\overline {\cal G}^{q,(1)}_3$ needed for $\Delta_{b}^{sv,(3)}|_\delta$ 
were not available. 
The recent results for $g^{b,2}_2$ and $g^{b,1}_3$ from \cite{Gehrmann:2014vha},
$\overline{{\cal G}}^{q,2}_2$ from \cite{deFlorian:2012za}
and the N$^3$LO $\overline{{\cal G}}^{q, (1)}_3$ 
from \cite{Ahmed:2014cla} can be used to obtain 
the missing $\delta(1-z)$ part namely $\Delta_{b}^{sv,(3)} |_{\delta}$.
This completes the evaluation of full threshold N$^3$LO contributions in QCD for
Higgs production through bottom anti-bottom annihilation at hadron colliders. 
Below we present our results up to N$^3$LO level after substituting $A_i,B_i,f_i,\gamma_i$ and $\beta_i$ 
terms as well as the constants $g^{b,i}_j$ from the form factors and $\overline{{\cal G}}^{b,(i)}_j$
from the soft distributions function in eqn. (\ref{ci3}).
%

%
\begin{align}
 \Delta_{b}^{sv,(3)} &= 
\delta (1-z) \Bigg\{ 
{C_A}^2 {C_F} \Big(\frac{13264}{315} {\zeta_2}^3 + \frac{2528}{27} {\zeta_2}^2 - \frac{1064}{3} {\zeta_2} {\zeta_3} 
    - 272 {\zeta_2} - \frac{400}{3} {\zeta_3}^2 - \frac{14212}{81} {\zeta_3} 
\nonumber \\
&
    - 84 {\zeta_5} + \frac{68990}{81}\Big)
+ {C_A} {C_F}^2 \Big( - \frac{20816}{315} {\zeta_2}^3 - \frac{62468}{135} {\zeta_2}^2 + \frac{27872}{9} {\zeta_2} {\zeta_3}
    + \frac{22106}{27} {\zeta_2} 
\nonumber \\
&
    + \frac{3280}{3} {\zeta_3}^2 - \frac{10940}{9} {\zeta_3} - \frac{37144}{9} {\zeta_5} - \frac{982}{3}\Big)
+ {C_A} {C_F} {n_f} \Big( - \frac{6728}{135} {\zeta_2}^2 + \frac{208}{3} {\zeta_2} {\zeta_3} 
\nonumber \\
&
    + \frac{3368}{81} {\zeta_2} 
    + \frac{2552}{81} {\zeta_3} - 8 {\zeta_5} - \frac{11540}{81}\Big)
+ {C_F}^3 \Big( - \frac{184736}{315} {\zeta_2}^3 + \frac{152}{5} {\zeta_2}^2 - 64 {\zeta_2} {\zeta_3} 
\nonumber \\
&
    - \frac{550}{3} {\zeta_2}
    + \frac{10336}{3} {\zeta_3}^2 - 1188 {\zeta_3} + 848 {\zeta_5} + \frac{1078}{3}\Big)
+ {C_F}^2 {n_f} \Big(\frac{12152}{135} {\zeta_2}^2 - \frac{5504}{9} {\zeta_2} {\zeta_3} 
\nonumber \\
&
    - \frac{2600}{27} {\zeta_2} + \frac{4088}{9} {\zeta_3} 
    + \frac{5536}{9} {\zeta_5} - \frac{70}{9}\Big)
+ {C_F} {n_f}^2 \Big(\frac{128}{27} {\zeta_2}^2 - \frac{32}{81} {\zeta_2} - \frac{1120}{81} {\zeta_3} + \frac{16}{27}\Big)
\Bigg\}
\nonumber \\
&
+ \D_0  \Bigg\{
{C_A}^2 {C_F} \Big(-\frac{2992}{15} {\zeta_2}^2 - \frac{352}{3} {\zeta_2} {\zeta_3} + \frac{98224}{81} {\zeta_2} + \frac{40144}{27} {\zeta_3}
    - 384 {\zeta_5} - \frac{594058}{729}\Big)
\nonumber \\
&
+ {C_A} {C_F}^2 \Big( \frac{1408}{3} {\zeta_2}^2 - 1472 {\zeta_2} {\zeta_3} + \frac{6592}{27} {\zeta_2} + \frac{32288}{9} {\zeta_3} 
    + \frac{6464}{27}\Big)
+ {C_A} {C_F} {n_f} \Big( \frac{736}{15} {\zeta_2}^2 
\nonumber \\
&
  - \frac{29392}{81} {\zeta_2} - \frac{2480}{9}  {\zeta_3} + \frac{125252}{729}\Big)
+ {C_F}^3 (-6144 {\zeta_2} {\zeta_3} - 1024 {\zeta_3} + 12288 {\zeta_5})
\nonumber \\
&
+ {C_F}^2 {n_f} \Big( - \frac{1472}{15} {\zeta_2}^2 - \frac{1504}{27} {\zeta_2} - \frac{5728}{9} {\zeta_3} + \frac{842}{9}\Big)
+ {C_F} {n_f}^2 \Big( \frac{640}{27} {\zeta_2} + \frac{320}{27} {\zeta_3} 
\nonumber \\
&
  - \frac{3712}{729}\Big)
\Bigg\}
+ \D_1  \Bigg\{
{C_A}^2 {C_F} \Big(\frac{704}{5} {\zeta_2}^2 - \frac{12032}{9} {\zeta_2} - 704 {\zeta_3} + \frac{124024}{81}\Big)
\nonumber \\
&
+ {C_A} {C_F}^2 \Big(\frac{3648}{5} {\zeta_2}^2 - \frac{20864}{9} {\zeta_2} - 5760 {\zeta_3} - \frac{544}{3}\Big)
+ {C_A} {C_F} {n_f} \Big(384 {\zeta_2} - \frac{32816}{81}\Big)
\nonumber \\
&
+ {C_F}^3 \Big(-\frac{14208}{5} {\zeta_2}^2 + 1024 {\zeta_2} - 960 {\zeta_3} + 256\Big)
+ {C_F}^2 {n_f} \Big( \frac{3200}{9} {\zeta_2} + 1280 {\zeta_3} - \frac{184}{3}\Big)
\nonumber \\
&
+ {C_F} {n_f}^2 \Big(\frac{1600}{81}-\frac{256}{9} {\zeta_2} \Big)
\Bigg\}
+ \D_2  \Bigg\{
{C_A}^2 {C_F} \Big(\frac{704}{3} {\zeta_2} - \frac{28480}{27}\Big)
+ {C_A} {C_F}^2 \Big(\frac{11264}{3} {\zeta_2} 
\nonumber \\
&
   + 1344 {\zeta_3} - \frac{10816}{9}\Big)
+ {C_A} {C_F} {n_f} \Big(\frac{9248}{27}-\frac{128}{3} {\zeta_2}\Big)
+ {C_F}^3 \Big( 10240 {\zeta_3} \Big) 
+ {C_F}^2 {n_f} \Big(\frac{1696}{9}
\nonumber \\
&
   -\frac{2048}{3} {\zeta_2} \Big) 
+  {C_F} {n_f}^2 \Big( - \frac{640}{27} \Big) 
\Bigg\}
+ \D_3  \Bigg\{
 {C_A}^2 {C_F} \Big( \frac{7744}{27} \Big) 
 + {C_A} {C_F}^2 \Big(\frac{17152}{9}-512 {\zeta_2}\Big)
\nonumber \\
&
 +  {C_A} {C_F} {n_f} \Big( -\frac{2816}{27} \Big)
 + {C_F}^3 (-3072 {\zeta_2} - 512) 
 + {C_F}^2 {n_f} \Big( -\frac{2560}{9} \Big) 
 + {C_F} {n_f}^2 \Big( \frac{256}{27} \Big)
\Bigg\}
\nonumber \\
&
+ \D_4  \Bigg\{
 {C_F}^2 {n_f} \Big( \frac{1280}{9} \Big) +  {C_A} {C_F}^2 \Big( - \frac{7040}{9} \Big)
\Bigg\}
+ \D_5  \Bigg\{
512 {C_F}^3
\Bigg\} \,.
\end{align}
 \begin{figure}[h]
 \centering
 \begin{minipage}[c]{0.48\textwidth}
\includegraphics[width=1.0\textwidth]{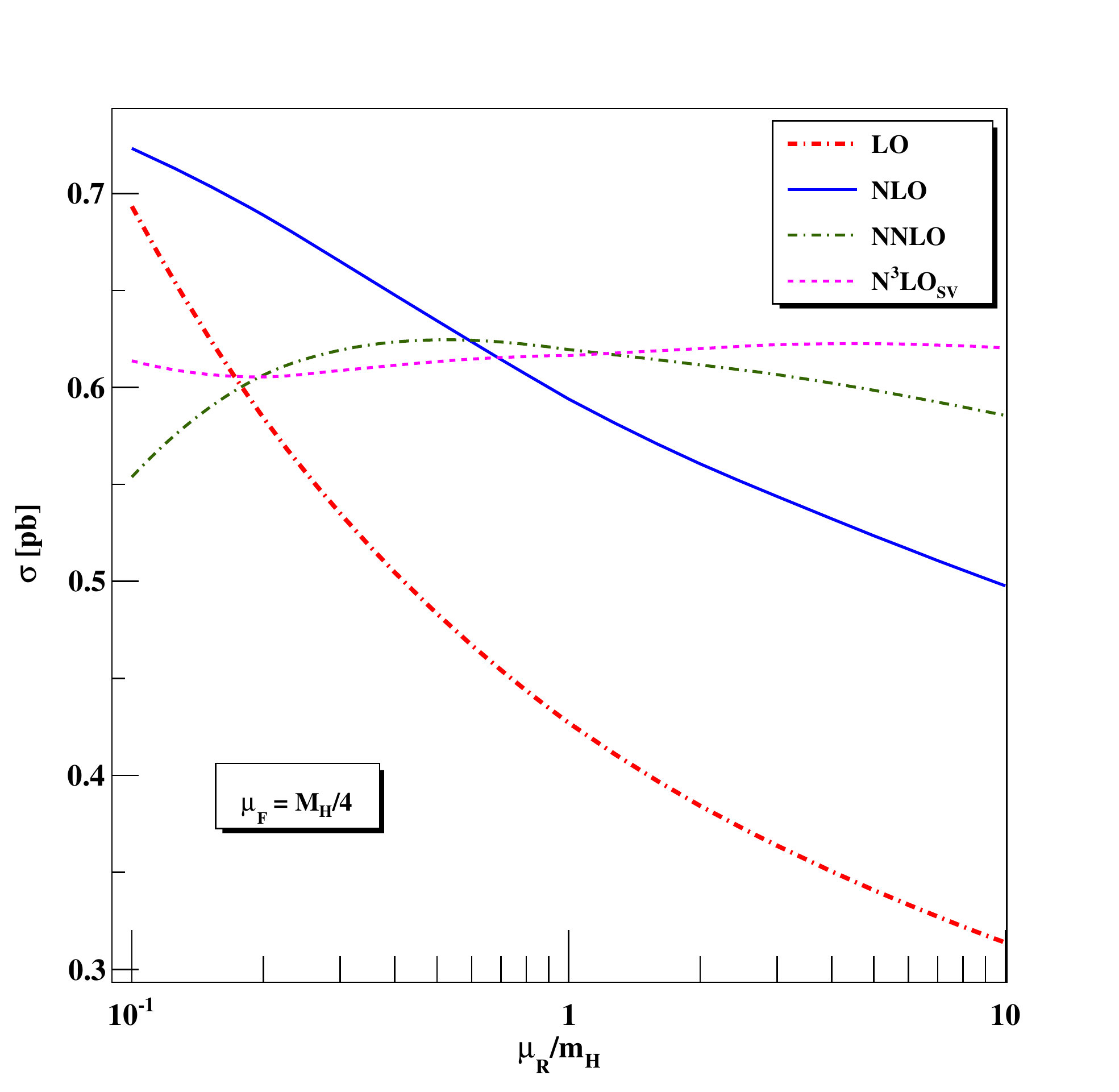}
\end{minipage}
\begin{minipage}[c]{0.48\textwidth}
\includegraphics[width=1.0\textwidth]{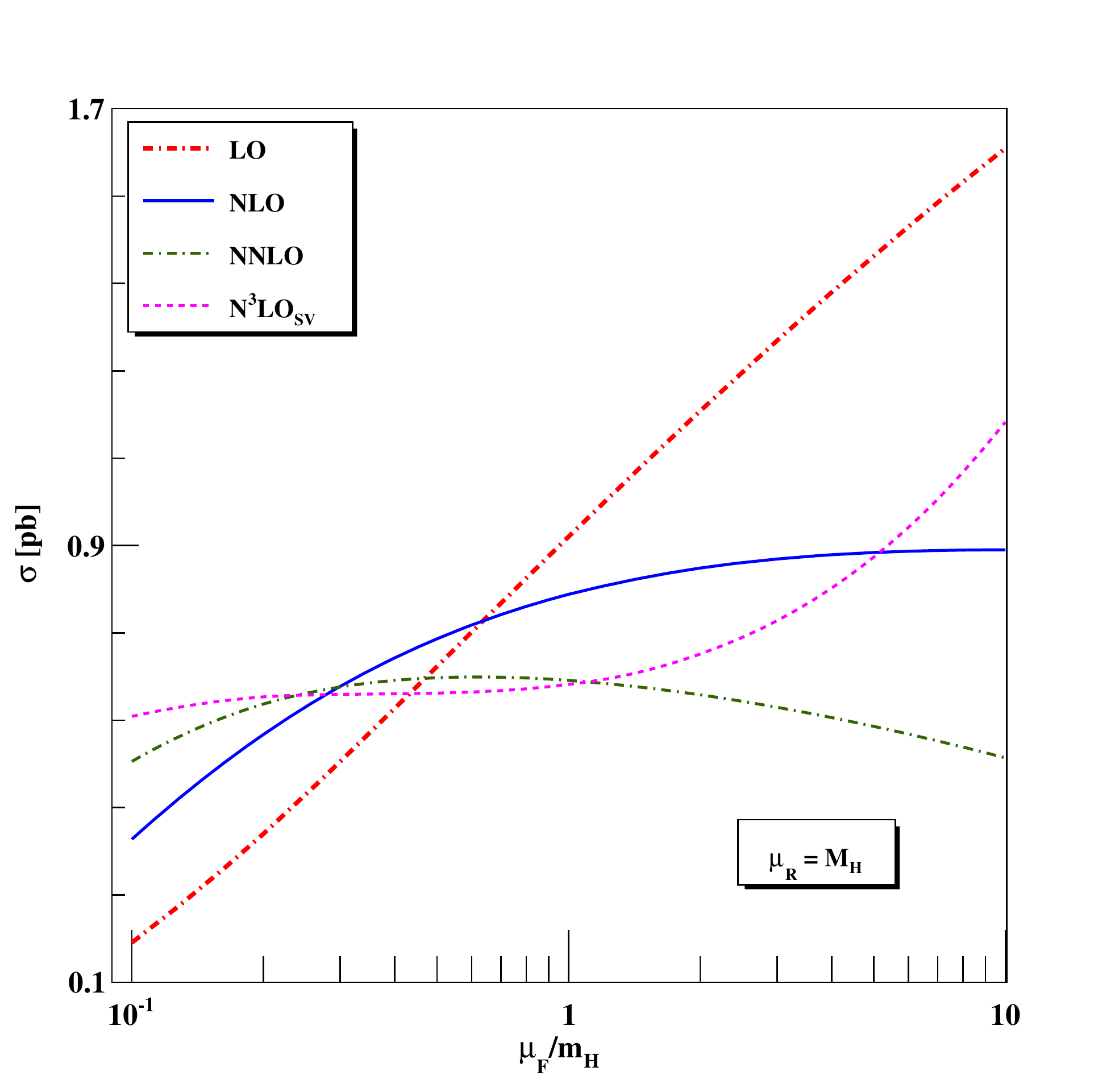}
\end{minipage}
\caption{\label{fig:murnmuf}
Total cross section for Higgs production in $b\bar{b}$ annihilation at various orders in $a_s$ as a function of $\mu_R/m_H$ (left panel) and of $\mu_F/m_H$ (right panel)
at the LHC with $\sqrt{s}=14$ TeV.
}
\end{figure}
The numerical impact of our results can be studied using the exact LO, NLO, NNLO
$\Delta^{b,(i)}_{ac},~ i=0,1,2$ and the threshold N$^3$LO result $\Delta^{sv,(3)}_{bb}$.
We have used $\sqrt{s} = 14$ TeV for the LHC, the $Z$ boson mass $M_Z=91.1876$ GeV 
and Higgs boson mass $m_H$ = 125.5 GeV throughout. 
The strong coupling constant $\alpha_s (\mu_R^2)$ is evolved 
using the 4-loop RG equations with 
$\alpha_s^{\text{N$^3$LO}} (m_Z ) = 0.117$ and for parton density sets we use 
MSTW 2008NNLO \cite{Martin:2009iq}.
The Yukawa coupling is evolved using 4 loop RG with $\lambda(m_b)=\sqrt{2} m_b(m_b)/\nu$ and $m_b(m_b)=4.3$ GeV.

The renormalization scale dependence is studied by varying $\mu_{R}$ between $0.1 ~ m_H$ and $10 ~ m_H$ 
keeping $\mu_{F}=m_{H}/4$ fixed. For the factorization scale, we have fixed $\mu_R=m_H$ and
varied $\mu_F$ between $0.1 ~ m_H$ and $10 ~ m_H$.  We find that the perturbation theory
behaves better if we include more and more higher order terms (see Fig.\ref{fig:murnmuf}).  

To summarize, we have systematically developed a framework to compute
threshold contributions in
QCD to the production of Higgs boson in bottom anti-bottom annihilation subprocesses at the 
hadron colliders.  Factorization of UV, soft and collinear singularities and  
exponentiation of their sum allow us to obtain threshold corrections   
order by order in perturbation theory.
Using the recently obtained N${}^3$LO soft distribution function for Drell-Yan production
and the three loop Higgs form factor with bottom anti-bottom quarks, we have obtained threshold
N${}^3$LO corrections to Higgs production through bottom anti-bottom annihilation.  
We have also studied the stability of our result under renormalization and factorization scales.

\section*{Acknowledgments}
We sincerely thank T. Gehrmann for constant encouragement and fruitful discussions.
We also thank M. K. Mandal for useful discussions. The work of
TA and NR has been partially supported by funding from RECAPP, Department of
Atomic Energy, Govt. of India.

\end{document}